\documentclass[prb,amsmath,amssymb, notitlepage, twocolumn,
nofootinbib,
superscriptaddress,
longbibliography
]{revtex4-2}

\usepackage{amsmath}
\usepackage{tabularx,graphicx}
\usepackage{epstopdf}
\usepackage{graphicx}
\usepackage{latexsym}
\usepackage{amssymb}
\usepackage{amsmath}
\usepackage{color, colortbl}
\usepackage{psfrag}
\usepackage{bbm}
\usepackage{bm}
\usepackage{titlesec}
\usepackage{dsfont}
\usepackage{feynmp}
\usepackage{slashed}
\usepackage{multirow}
\usepackage{url}
\usepackage{natbib}
\usepackage{soul}
\usepackage{lipsum}

\newcommand{\beq}{\begin{eqnarray}}
	\newcommand{\eeq}{\end{eqnarray}}

\newcommand{\bsp}{\begin{split}}
	\newcommand{\esp}{\end{split}}

\usepackage{color}
\definecolor{darkblue}{rgb}{0.,0.,0.4}
\definecolor{darkred}{rgb}{0.5,0.,0.}
\definecolor{BlueViolet}{RGB}{138,43,226}
\definecolor{SkyBlue}{RGB}{30,144,255}
\definecolor{DarkGreen}{RGB}{0,100,0}
\usepackage[pdftex,colorlinks=true,linkcolor=darkblue,citecolor=blue,urlcolor=darkred]{hyperref}
\usepackage[normalem]{ulem}

\usepackage{mathrsfs}
\usepackage{comment}

\begin{document}
	
	\title{
		Flat band excitons and quantum metric
	}
	
	\author{Xuzhe Ying}
	\affiliation{Department of Physics, Hong Kong University of Science and Technology,
		Clear Water Bay, Hong Kong SAR, China}
		
	\author{K.T. Law}
	\affiliation{Department of Physics, Hong Kong University of Science and Technology,
		Clear Water Bay, Hong Kong SAR, China}
	
	\begin{abstract}
		
		We discuss the excitons in flat band systems. Quantum metric plays a central role in determining the properties of single exciton excitation as well as the exciton condensate. While the electrons and holes are extremely heavy in flat bands, the excitons (boundstate of an electron-hole pair) could be light and mobile. In particular, we show that the inverse of exciton's effective mass tensor is proportional to the quantum metric tensor. Meanwhile, the flat band excitons have a finite size, lower bounded by the trace of the quantum metric tensor. Given the properties of single exciton excitation, one can argue for the formation of exciton condensate. Because of the quantum metric, the exciton condensate can support dissipationless counterflow supercurrent, implying the stability of exciton condensate.
		
	\end{abstract}
	
	\maketitle
	
	%\tableofcontents

	\section{Introduction} \label{sec: intro}
	
	The discovery and study of various strongly correlated phases in flat band system \cite{cao2018unconventional,cao2018correlated,sharpe2019emergent,serlin2020intrinsic,cao2021nematicity,gu2022dipolar,xu2020correlated,liu2022crossover,xie2021fractional,cai2023signatures,park2023observation,PhysRevLett.132.036501}, in particular Moire materials, brings the concept of quantum geometric tensor to the central stage \cite{vanderbilt2018berry,provost1980riemannian,cheng2010quantum,PhysRevB.90.165139,PhysRevLett.114.236802,PhysRevB.96.165150,jackson2015geometric,PhysRevResearch.2.023237}. Even though the electron's kinetic energy is quenched in flat band systems, the other aspect of the electron's kinetics, namely the Bloch wavefunction, can be highly nontrivial \cite{bernevig2013topological,shen2012topological,RevModPhys.82.3045}. Quantum geometric tensor is one such measure of the ``quantum kinetics'' encoded in the Bloch wavefunction. Remarkably, the imaginary part of the quantum geometric tensor is the famous Berry curvature, which shows up explicitly in the semiclassical kinetic equation of Bloch electrons \cite{PhysRevB.59.14915,PhysRevB.77.035110,RevModPhys.82.1959,kamenev2023field,vanderbilt2018berry}. Even more interestingly, the integration of Berry curvature over Brillouin zone defines a topological invariant, known as the Chern number, which aids classifying insulating phases of matter \cite{bernevig2013topological,shen2012topological,RevModPhys.82.3045}. 
	
	At the meantime, the real part of the quantum geometric tensor is usually referred to as the Fubini-Study metric between quantum states (or quantum metric for short). The quantum metric is a mathematical measure of the difference between two quantum states which are parametrically close to each other \cite{provost1980riemannian,cheng2010quantum}. Remarkably, the quantum metric is closely related to many physical properties of solids, including the size of maximally localized Wannier functions \cite{PhysRevB.56.12847}, superfluid stiffness for flat band superconductors \cite{PhysRevLett.127.170404,tian2023evidence}, and sum rules \cite{mao2023diamagnetic,kruchkov2023spectral,onishi2024quantum}. In this work, we focus on the quantum metric effects on flat band excitons.
	
	Significant attention has been put on the impact of Berry curvature on the excitons and exciton condensation. Conventionally, an exciton is conceived as a hydrogen-like boundstate formed by an electron and a hole in solid \cite{kittel2005introduction,glutsch2004excitons}. Recently, various non-hydrogenic features of excitons were discovered \cite{PhysRevLett.126.137601,PhysRevLett.115.166802,PhysRevLett.115.166803}. Examples include nontrivial exciton band topology \cite{PhysRevLett.126.137601} and the Berry curvature induced energy splitting of exciton states with different angular momentum \cite{PhysRevLett.115.166802,PhysRevLett.115.166803}.
	
	Excitons, as (non-ideal) composite bosons, are susceptible to condense to form a superfluid phase \cite{keldysh2017coherent}. Exciton condensate was observed in systems of quantum Hall bilayer \cite{eisenstein2004bose,doi:10.1146/annurev-conmatphys-031113-133832,PhysRevLett.84.5808,PhysRevLett.88.126804,PhysRevLett.88.126804,su2008make,nandi2012exciton}. Due to the topological nature of quantum Hall systems, the exciton condensate bears a composite Fermi liquid (cFL) description \cite{PhysRevB.95.085135}. Exciton condensation was identified as the Fermi surface instability of the cFL, even though the Landau levels are non-dispersive. Numerous interesting excitonic phases of matter were proposed to study the interplay of interaction and topology \cite{PhysRevLett.121.126601,PhysRevB.105.235121,PhysRevX.13.031023}.
	
	In this article, we point out that the quantum metric (real part of the quantum geometric tensor) could also significantly influence the exciton properties, in addition to the Berry curvature and topology. The role of quantum metric is most profound in flat band systems, for both single exciton excitation and exciton condensate. Indeed, we will show that the exciton's size and kinetic energy are very sensitive to the quantum metric of the underlying Bloch bands, in which the electron and hole resides. Meanwhile, regarding the exciton condensate, the quantum metric gaurantees the stability of it. The stability of exciton condensate manifests itself in the possibility of a dissipationless counterflow (CF) supercurrent. The reason behind is deeply rooted in the fact that quantum metric is a measure of quantum fluactuation in space \cite{PhysRevB.56.12847,PhysRevB.62.1666,onishi2024quantum}. 
	
	To demonstrate the point above, we consider a hopping model defined on the Lieb lattice, which supports an isolated non-topological flat band. The advantage of Lieb lattice model is that the flat band shows widely tunable quantum metric. We consider a bilayer system of Lieb lattice hopping model, where chemical potential is close to the flat bands. We first study an exciton formed by an electron and a hole, residing in the flat bands of the two layers separately. We find that the exciton's effective mass tensor is proportional to the quantum metric, with the proportionality being the interaction strength. In addition, the exciton's wavefunction has finite spatial spread. The quadratic spread is shown to be twice of the quantum metric. Meanwhile, deep in the superfluid phase, a non-dissipative CF supercurrent is possible to establish in the bilayer system. The reason behind is still the quantum metric. Therefore, one can conclude that quantum metric stablizes the superfluid phase.
	
	We should mention that our study on the exciton's effective mass and superfluid supercurrent are very similar to Ref.~\cite{PhysRevB.98.220511} and Ref.~\cite{PhysRevB.101.060505}, respectively. Nevertheless, we point out another important piece of information - the exciton size - which is more or less overlooked. Indeed, previous studies revealed that the exciton size effect may affect the effective exciton-exciton interaction as well as the formation of exciton condensate in conventional bilayer system \cite{PhysRevB.84.075130}. Now, in flat band, the conventional Bohr radius vanishes due to the infinite electronic effective mass. Instead, quantum metric determines a length scale for spatial quantum fluctuations. Qualitatively, spatial fluctuation of particles makes a phase coherent state possible.
	
	The rest of the manuscript is organized as follows. In Sec.~\ref{sec: EX&LL}, we study the excitons in quantum Hall bilayers, where we argue that magnetic length may be understood as the quantum metric of Landau level wavefunctions. Sec.~\ref{sec:LiebLattice&Excitons} is devoted to the study of effective mass and size of excitons in the \emph{non-topological flat} band in Lieb lattice hopping model. Sec.~\ref{sec:EXCondensate} discusses the implications on the transport and stability of the exciton condensation. Sec.~\ref{sec: discussion} concludes this article and addresses a few open questions. 
	
	This article is further supported by an appendix on certain technical details. Appendix.~\ref{sec:LLEX_RealSpaceWF_ExplicitForm} provides details on the real space wavefunction of excitons. Appendix.~\ref{sec:SchrodingerEqaution&GaugeInvariance} discusses the gauge properties of the Schrodinger equation and construction of a few special exciton solutions. Appendix.~\ref{sec: MF_EXCondensate} derives the expression for CF supercurrent.
	
	\section{Excitons in bilayer quantum Hall systems} \label{sec: EX&LL}
	
	\begin{figure}[t]
		\centering
		\includegraphics[scale = 1]{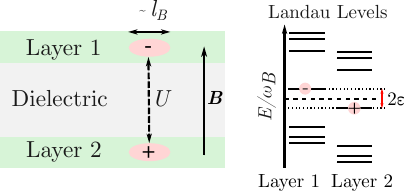}
		\caption{Schematic quantum Hall bilayer system: Left: two layers of (Dirac) material are separated by a dielectric spacer. An out of plane magnetic field $\boldsymbol{B}$ is applied. The electron and hole may form a bound state by the presence of attactive interaction $U$. The wavefunctions in Landau level (LL) have finite spatial extensions, leading to the exciton size on the order of magnetic length $l_{B}$. Right: Schematic of Landau levels (LL). A gate voltage is assumed such that the zeroth LLs of the two layers have a small energy off-set $2\varepsilon$. An exciton is formed by an electron in the zeroth LL of layer 1 and a hole in the zeroth LL of layer 2.}
		\label{fig:Exciton&QHBilayer}
	\end{figure}
	
	In this section, we study the properties of a single exciton excitation in quantum Hall bilayer system. From semiclassical picture of electronic orbits in magnetic field, one would argue that the size of excitons is determined by the magnetic length. Here, we confirm this semiclassical expectation by solving the quantum mechanical excitonic wavefunction.
	
	In particular, we point out that the magnetic length could be understood as the quantum metric of the Landau level wavefunctions. The notion of quantum metric is readily generalizable to many other flat band systems. Indeed, in the later sections, we will show the relation of exciton and exciton condensate with quantum metric in a non-topological flat band system.
	
	As an example, we consider a Dirac material based bilayer system Fig.~\ref{fig:Exciton&QHBilayer}. With an out-of-plane magnetic field $B$, the Landau levels within each layer can be obtained from the Hamiltonian $H_0=-i\partial_x\sigma_x+(-i\partial_y-Bx)\sigma_y$, where $\sigma_{x,y}$ are the Pauli matrices in sublattice/orbital basis. Here, Landau gauge $\boldsymbol{A}=(0,Bx)$ is assumed. We focus on the zeroth Landau level (LL). The wavefunction of zeroth Landau level is given by:
	\begin{equation}
		\begin{split}
			\psi_{k_y}(\boldsymbol{r})=\begin{bmatrix}
				\psi^{\text{A}}_{k_y}(\boldsymbol{r}) \\ \psi^{\text{B}}_{k_y}(\boldsymbol{r})
			\end{bmatrix}
			=e^{ik_yy} \left(\frac{B}{\pi}\right)^{\frac{1}{4}}e^{-\frac{B}{2}\left(x-\frac{k_y}{B}\right)^2}\begin{bmatrix}
				1 \\ 0
			\end{bmatrix}
		\end{split}
		\label{Eq:0th_LL_WF}
	\end{equation}
	where $k_y$ is the momentum in $y$-direction. Before proceeding to construct exciton wavefunction, we point out that the quantum metric of Landau level wavefunction defines a length scale, which is proportional to the magnetic length.
	
	To define the quantum metric, we take the momentum $k_y$ as a parameter and consider the following wavefunction overlap:
	\begin{equation}
		1- \left|\int dx\psi_{k_y}^{\dagger}(x,y)\psi_{k_y+q_y}(x,y)e^{-iq_yy}\right|\approx \frac{1}{2}gq_y^2
		\label{Eq:QD_LL}
	\end{equation}
	where the left hand side is usually referred to as the quantum distance between zeroth LL wavefunctions at different momenta $k_y$ and $k_y+q_y$ \cite{provost1980riemannian,cheng2010quantum}. Namely, when $q_y=0$, the two wavefunction are identical. Therefore, the quantum distance is zero. For small $q_y\ll B^{\frac{1}{2}}$, the two wavefunctions in Eq.~(\ref{Eq:QD_LL}) is slightly different, namely a small quantum distance $\frac{1}{2}gq_y^2$. The quantum metric is given by the quantity $g$. Direct calculation shows that the quantum metric is proportional to magnetic length squared:
	\begin{equation}
	 	g=\frac{1}{2B}.
	\end{equation}
	Indeed, the square root of the quantum metric defines a basic length scale in flat band systems.
	
	Below, we show that the size of an exciton in quantum Hall bilayer system is also given by the magnetic length (by coincidence or not). In later section, we will point out that this is no coincidence. Indeed, the size of excitons in flat bands will be shown to be lower bounded by quantum metric length $\gtrsim\sqrt{g}$. 
	
	With a contact interaction, we are able to find the exact exciton wavefunction as follows:
	\begin{equation}
		|e_1,h_2;\boldsymbol{Q} \rangle = \mathcal{N}\int\frac{dk_y}{2\pi}e^{ik_y\frac{Q_x}{B}}c^{\dagger}_{k_y+Q_y}d_{k_y}|\text{G.S.}\rangle
		\label{Eq:Ex_LLBilayer}
	\end{equation}
	with $\mathcal{N}$ being the normalization constant. A few comments are due. 
	
	First, to construct the exciton wavefunction above, an insulating groundstate $|\text{G.S.}\rangle$ is assumed, where the zeroth LL in layer 1 (2) is fully empty (filled), see Fig.~\ref{fig:Exciton&QHBilayer}. On top of the groundstate, an electron and a hole living in the zeroth Landau level are created by the operators $c^{\dagger}_{k_y+Q_y}$ and $d_{k_y}$, respectively. Such electron-hole pair carries a total momentum $Q_y$ in $y$-direction. With proper superposition coefficients, Eq.~(\ref{Eq:Ex_LLBilayer}) describes an exciton, namely a boundstate of electron and hole, with total momentum $\boldsymbol{Q}=(Q_x, Q_y)$. Notice that due to the choice of Landau gauge, the wavefunction of a single electron or hole does not have translation invariance in $x$-direction. Therefore, momentum in $x$-direction is not a good quantum number for single particle state. However, the Schrodinger equation for excitons is translation invariant in both directions. In particular, the translation invariance in $x$-direction is dictated by the invariance under the shift $k_y\rightarrow k_y+\delta k_y$. Therefore, the exciton wavefunction is labeled by total momentum $\boldsymbol{Q}$.
	
	Second, interaction is important here, given the fact that the kinetic energy is quenched. For simplicity, we consider a contact interaction:
	\begin{equation}
		H_{\text{int}}=U\int d^2r\sum_{\alpha=\text{A,B}}c^{\dagger}_{\alpha}(\boldsymbol{r})d^{\dagger}_{\alpha}(\boldsymbol{r})d_{\alpha}(\boldsymbol{r})c_{\alpha}(\boldsymbol{r})
		\label{Eq:Int_LL_Bilayer}
	\end{equation}
	where $c(d)^{\dagger}_{\alpha}(\boldsymbol{r})$ is the fermion creation operator in layer 1(2) that creates a fermion at location $\boldsymbol{r}$ and orbital $\alpha$; $c_{\alpha}(\boldsymbol{r})$ and $d_{\alpha}(\boldsymbol{r})$ are the corresponding annihilation operators. The fermion creation and annihilation operators are projected onto the lowest Landau level \cite{PhysRevB.103.205413}:
	\begin{equation}
		\begin{split}
			&c_{A}(\boldsymbol{r})=\int\frac{dk_y}{2\pi}\psi^{\text{A}}_{k_y}(\boldsymbol{r})c_{k_y},\ \ \ c_B(\boldsymbol{r})=0,
		\end{split}
		\label{Eq:OPProjection_0th_LL}
	\end{equation}
	and similarly for $c^{\dagger}_{\alpha}(\boldsymbol{r})$, $d_{\alpha}(\boldsymbol{r})$, and $d^{\dagger}_{\alpha}(\boldsymbol{r})$. With the projected interaction Hamiltonian, we are able to write down the Schrodinger equation for excitons as $(E-2\epsilon)|e_1,h_2;\boldsymbol{Q} \rangle=H_{\text{int}}|e_1,h_2;\boldsymbol{Q} \rangle$.
	
	It turns out that the exact exciton wavefunction is given by Eq.~(\ref{Eq:Ex_LLBilayer}), with a binding energy given by:
	\begin{equation}
		E_{\text{b}}(\boldsymbol{Q})=\left|E-E_{GS}-2\epsilon-\frac{1}{4}UB\right|=\frac{1}{4}UBe^{-\frac{\boldsymbol{Q}^2}{2B}}
	\end{equation}
	One important feature is that the excitons acquire a finite effective mass, $\frac{1}{M}=\frac{UB}{4B}\propto E_{\text{int}}g$. Here, we interpret the inverse effective mass as being proportional to the interaction energy, $E_{\text{int}}\sim UB$, and quantum metric $g\sim\frac{1}{B}$. At this point, such interpretation seems arbitrary. Nevertheless, the interpretation of $\frac{1}{M}\propto E_{\text{int}}g$ can be generalized to other flat band systems. 
	
	Another important aspect of the exciton wavefunction, Eq.~(\ref{Eq:Ex_LLBilayer}), is the spatial extension of the wavefunction. By inversing the relation of Eq.~(\ref{Eq:OPProjection_0th_LL}), we are able to rewrite the exciton wavefunction in terms of the operators in real space:
	\begin{equation}
		\begin{split}
			&|e_1,h_2;\boldsymbol{Q} \rangle =\int d^2r_1 d^2r_2\psi^{\boldsymbol{Q}}_{\text{AA}}(\boldsymbol{r}_1,\boldsymbol{r}_2)c^{\dagger}_{\text{A}}(\boldsymbol{r}_1)d_{\text{A}}(\boldsymbol{r}_2)|\text{G.S.}\rangle
		\end{split}
		\label{Eq:LLEX_RealSpaceWF}
	\end{equation}
	with $\psi^{\boldsymbol{Q}}_{\text{AA}}(\boldsymbol{r}_1,\boldsymbol{r}_2)=\mathcal{N}\int\frac{dk_y}{2\pi}e^{ik_y\frac{Q_x}{B}}\left[\psi^{\text{A}}_{k_y+Q_y}(\boldsymbol{r}_1)\right]^*\psi^{\text{A}}_{k_y}(\boldsymbol{r}_2)$ is the real space exciton wavefunction. The explicit expression can be found in Appendix.~\ref{sec:LLEX_RealSpaceWF_ExplicitForm}.
	
	The exciton size can be defined by the separation between the electron and hole, $\Delta\boldsymbol{r}=\boldsymbol{r}_1-\boldsymbol{r}_2$, through the real space wavefunction $\psi^{\boldsymbol{Q}}_{\text{AA}}(\boldsymbol{r}_1,\boldsymbol{r}_2)$. There are many measures of the separation. For example, the average separation is given by $\langle\Delta\boldsymbol{r}\rangle=\left(\frac{Q_y}{B},\frac{Q_x}{B}\right)$. For small $|\boldsymbol{Q}|\ll\sqrt{B}$, the average separation between electron and hole is much smaller than magnetic length. In this case, it's better to define the exciton size through the variance in the separation. We find that the variance in the separation between electron and hole in the exciton wavefunction is given by:
	\begin{equation}
		\langle \left(\Delta x - \frac{Q_y}{B}\right)^2\rangle = \frac{1}{B}, \ \ \ \ \langle \left(\Delta y-\frac{Q_x}{B}\right)^2\rangle  = \frac{1}{B}
	\end{equation}
	which is exactly magnetic length squared, $l_B^2=\frac{1}{B}$. At the mean time, the exciton size (squared) as derived above happen to be twice of the quantum metric, $g$. Below, we study a non-topological flat band and show that this is no coincidence.
	
	Before ending this section, we comment on the exciton condensate in quantum Hall bilayers. From the analysis above, we see that excitons in quantum Hall bilayer have finite kinetic energy and spatially spreading wavefunctions. The latter fact will affect interaction energy between excitons \cite{PhysRevB.84.075130}. Hence, it is possible to find a regime of exciton density where kinetic energy dominates, leading to a condensation of excitons \cite{eisenstein2004bose,doi:10.1146/annurev-conmatphys-031113-133832,liu2022crossover}. A more accurate description is based on the composite Fermi liquid theory, where an interplay of interaction and topology becomes important \cite{PhysRevB.95.085135}. In the rest of the manuscript, we turn our attention to \emph{non-topological} flat band and excitons within them. We will show that quantum metric plays an important role.

	\section{Flatband models and excitons} \label{sec:LiebLattice&Excitons}
	
	In this section, we study the exciton properties in a \emph{non-topological} flat band system. We focus on a particular example of Lieb lattice model. The flat band in Lieb lattice hopping model is not topological. Namely, the Chern number vanishes. However, the quantum metric (real part of the quantum geometric tensor) is nonzero and is highly tunable. Below, we will introduce the details of the Lieb lattice model and then study the spatially indirect excitons formed in the system of Lieb lattice bilayer. In particular, we will point out that even though the electron and holes are extremely heavy in flat bands. the composite particle, namely the exciton, acquires a finite effective mass and thus mobile. In addition, the excitons also acquire a finite size. The mechanism behind is closely related to the quantum metric.
	
	\subsection{Lieb lattice hopping model}
	
	\begin{figure}[t]
		\centering
		\includegraphics[scale = 1.25]{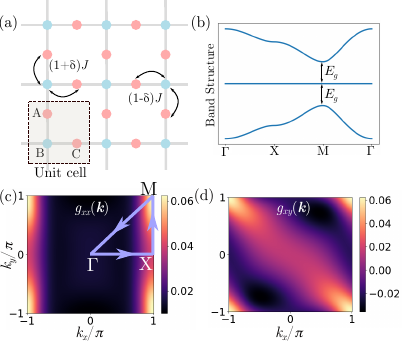}
		\caption{Lieb Lattice: (a) Lieb lattice is a square lattice. Each unit cell (gray box) contains three orbitals. The hopping amplitudes are $(1\pm\delta)J$ for intra-unit-cell and inter-unit-cell hoppings respectively. (b) Electronic band structure: There is a flat band in the middle of the spectrum, separated from two dispersive bands by an energy gap $E_{\text{g}}\sim \delta J$. The band structure is plotted along a contour (purple) indicated in (c). (c-d) Quantum metric $g_{xx}(\boldsymbol{k})$ and $g_{xy}(\boldsymbol{k})$ in Brillouin zone. The distribution of quantum metric in Brillouin zone is not uniform.}
		\label{fig:LiebLattice_LatticeStructure&BandStructure&QM}
	\end{figure}
	
	As shown in Fig.~\ref{fig:LiebLattice_LatticeStructure&BandStructure&QM}(a), Lieb lattice model is a tight binding model defined on square lattice. There are three sublattice within a unit cell. The sublattices are labeled as A, B, and C. To be specific, the choice of bulk unit cell is indicated by the gray box in Fig.~\ref{fig:LiebLattice_LatticeStructure&BandStructure&QM}(a). The hopping parameter is $t_{\text{intra}}=(1+\delta)J$ and $t_{\text{inter}}=(1-\delta)J$ for intra-unit-cell and inter-unit-cell hopping parameters respectively. The Bloch Hamiltonian of the Lieb lattice hopping model is given by:
	\begin{equation}
		H(\boldsymbol{k})=\begin{bmatrix}
			0 & a_{\boldsymbol{k}} & 0\\
			a_{\boldsymbol{k}}^* & 0 & b_{\boldsymbol{k}}\\
			0 & b_{\boldsymbol{k}}^* & 0
		\end{bmatrix}
		\label{Eq:LiebLattice_BlochH}
	\end{equation}
	with $a_{\boldsymbol{k}}=-J(1+\delta)-J(1-\delta)e^{ik_y}$ and $b_{\boldsymbol{k}}=-J(1+\delta)-J(1-\delta)e^{ik_x}$.
	
	 Fig.~\ref{fig:LiebLattice_LatticeStructure&BandStructure&QM}(b) shows the electronic band structure of a single layer of Lieb lattice hopping model. There is an \emph{isolated} flat/dispersionless band separated from two dispersive bands. The energy gap between the flat band and the dispersive bands is given by $E_g=2\sqrt{2}\delta J$.
	 
	 The Bloch wavefunction for the flat band is given by:
	 \begin{equation}
	 	\psi_{\boldsymbol{k}}(\boldsymbol{r})=e^{i\boldsymbol{k}\cdot\boldsymbol{r}}u_{\boldsymbol{k}},\ u_{\boldsymbol{k}}=\mathcal{N}_{\boldsymbol{k}}\left[-b_{\boldsymbol{k}},0,a_{\boldsymbol{k}}^*\right]^{\text{T}}
	 	\label{Eq:FB_BlochWF}
	 \end{equation}
	with normalization constant is $\mathcal{N}_{\boldsymbol{k}}=\sqrt{\left|a_{\boldsymbol{k}}\right|^2+\left|b_{\boldsymbol{k}}\right|^2}$. The Chern number of the flat band vanishes. Nevertheless, this flat band carries non-vanishing quantum metric. The quantum metric can be defined through the quantum distance between the periodic part of the Bloch wavefunctions, i.e., $1-\left|u^{\dagger}_{\boldsymbol{k}}u_{\boldsymbol{k}+\boldsymbol{q}}\right|\approx \frac{1}{2}g_{\mu\nu}(\boldsymbol{k})q^{\mu}q^{\nu}$ for small momentum difference $|\boldsymbol{q}|\ll 1$. More explicitly, the quantum metric can be computed as \cite{provost1980riemannian,cheng2010quantum,PhysRevB.56.12847}:
	\begin{equation}
		g_{\mu\nu}(\boldsymbol{k})=\text{Re} \left[\left(\partial_{\mu}u^{\dagger}_{\boldsymbol{k}}\right)\left(1-u_{\boldsymbol{k}}u^{\dagger}_{\boldsymbol{k}}\right)\left(\partial_{\nu}u_{\boldsymbol{k}}\right)\right]
	\end{equation}
	where $\partial_{\mu}=\partial_{k^{\mu}}$. Fig.~\ref{fig:LiebLattice_LatticeStructure&BandStructure&QM}(c)-(d) plots two componets of the quantum metric, $g_{xx}(\boldsymbol{k})$ and $g_{xy}(\boldsymbol{k})$, in Brillouin zone. The distribution of quantum metric is apparently not uniform in B.Z..  Nevertheless, a useful quantity is the averaged quantum metric over the Brillouin zone (B.Z.):
	\begin{equation}
		g_{\mu\nu}=\frac{(2\pi)^2}{\mathcal{A}_{\text{B.Z.}}}\int_{\text{B.Z.}}\frac{d^2k}{(2\pi)^2}g_{\mu\nu}(\boldsymbol{k})
	\end{equation}
	where $\mathcal{A}_{\text{B.Z.}}$ is the area of the Brillouin zone. In particular, the trace, $\text{tr}(g_{\mu\nu})=g_{xx}+g_{yy}$, defines the square of a fundamental length scale related to the isolated flat band. It determines the minimal spread of Wannier functions \cite{PhysRevB.56.12847}. As will be shown below, the length scale as defined from $\text{tr}(g_{\mu\nu})$ plays an analogous role of magnetic length in quantum Hall bilayer.

	 Before ending, we should comment on the tunability of quantum metric, in particular, $\text{tr}g_{\mu\nu}$. For the Lieb lattice hopping model under consideration, the quantum metric can be tuned solely by the parameter $\delta\in(0,1)$. When $\delta\rightarrow0$, quantum metric, $\text{tr}g_{\mu\nu}$, diverges. This is because all the sites are in almost resonance (connected by hopping with similar amplitudes). Quantum metric decreases with increasing $\delta$ and vanishes at $\delta=1$. Indeed, at $\delta=1$, the lattice becomes disconnected so that minimal Wannier wavefunction is fully localized within a unit cell, see Fig.~\ref{fig:LiebLattice_LatticeStructure&BandStructure&QM}(a).
	
	\subsection{Excitons in Lieb lattice bilayer}
	
	\begin{figure}[t]
		\centering
		\includegraphics[scale = 1.25]{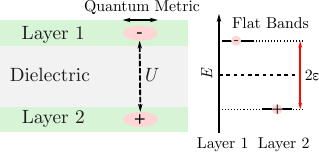}
		\caption{Schematic of Lieb lattice bilayer and flat band exciton. Left: two layers of Lieb lattices are separated by a dielectric spacer. The electron and hole may form a bound state by the presence of attactive interaction $U$. The electronic wavefunctions have finite spatial extensions, lower bounded by quantum metric. Accordingly, excitons also have a size bounded by quantum metric. Right: A gate voltage is assumed such that the flat bands of the two layers have a small energy off-set $2\varepsilon$. An exciton is formed by an electron in the flat band of layer 1 and a hole in the flat band of layer 2. The dispersive bands are assumed to be extremely far away in energy.}
		\label{fig:LiebLattice_Exciton_Setup}
	\end{figure}
	
	In this subsection, we describe the process to construct an exciton wavefunction.
	
	Fig.~\ref{fig:LiebLattice_Exciton_Setup} is a schematic plot of Lieb lattice bilayer system for studying spatially indirect flat band excitons. The backgates (which is not drawn explicitly) introduce a small energy offset $2\varepsilon$ between the flat bands of the two layers, such that the flat band in layer 1 is empty while the flat band in layer 2 is filled. Meanwhile, the dispersive bands are assumed to be very far away, namely $E_g\sim\delta J\gg \varepsilon$. The assumption of the dispersive bands being very far away exaggerates the role of flat bands in forming excitons. Given that the band structure is basically featureless for flat band, the properties of excitons are mainly determined by the Bloch wavefunction of the flat bands. Special attention is put on the quantum metric.
	
	To study an exciton, we consider an electron-hole pair in the bilayer system as shown in Fig.~\ref{fig:LiebLattice_Exciton_Setup}. For instance, we consider a state of the form
	 \begin{equation}
	 	|e_1,h_2;\boldsymbol{Q}\rangle = \int\frac{d^2k}{(2\pi)^2}\psi(\boldsymbol{k},\boldsymbol{Q})c^{\dagger}_{\boldsymbol{k}+\boldsymbol{Q}}d_{\boldsymbol{k}}|\text{G.S.}\rangle
	 	\label{Eq:GeneralP-HState}
	 \end{equation}	
	 where $|\text{G.S.}\rangle$ is the insulating groundstate with empty flat band in layer 1 and filled flat band in layer 2, Fig.~\ref{fig:LiebLattice_Exciton_Setup}; $c^{\dagger}_{\boldsymbol{k}}$ is the electron creation operator with momentum $\boldsymbol{k}$ in the flat band of layer 1; $d_{\boldsymbol{k}}$ is the electron annihilation (thus hole creation) operator with momentum $\boldsymbol{k}$ in the flat band of layer 2. Therefore, the state $|e_1,h_2;\boldsymbol{Q}\rangle $ describes a superposition state of electron-hole pairs with total momentum $\boldsymbol{Q}$, with superposition coefficients given by the function $\psi(\boldsymbol{k},\boldsymbol{Q})$. From Schrodinger equation, we will be able to determine the energy of such electron-hole pairs as well as the wavefunctions, namely the function $\psi(\boldsymbol{k},\boldsymbol{Q})$.
	 
	For flat bands, the kinetic energy is quenched. As a consequence, interaction is important. For simplicity, we consider a contact interaction:
	\begin{equation}
		H_{\text{int}}=U\sum_{\boldsymbol{r}}\sum_{\alpha=\text{A,B,C}}c^{\dagger}_{\alpha}(\boldsymbol{r})d^{\dagger}_{\alpha}(\boldsymbol{r})d_{\alpha}(\boldsymbol{r})c_{\alpha}(\boldsymbol{r})
		\label{Eq:ContactInteraction}
	\end{equation}
	where $\boldsymbol{r}$ labels the location of the unit cell; $\alpha=\text{A,B,C}$ labels the sublattices within a unit cell. Here, the electron creation/annihilation operators are written in real space. The fermion operators are projected onto the flat band \cite{PhysRevB.103.205413}:
	\begin{equation}
		c_{\alpha}(\boldsymbol{r})=\int_{\text{B.Z.}}\frac{d^2k}{(2\pi)^2}\psi_{\boldsymbol{k},\alpha}(\boldsymbol{r})c_{\boldsymbol{k}}
	\end{equation}
	and similarly for other fermionic operators. Here $\psi_{\boldsymbol{k},\alpha}$ is the $\alpha$-th component of the Bloch wavefunction at momentum $\boldsymbol{k}$ in flat band, Eq.~(\ref{Eq:FB_BlochWF}).
	
	We should comment on a few approximations. First, we assume interaction $U>0$, which corresponds to attractive interaction between electron and hole. Second, we assume the interaction strength is relatively weak, namely $U\ll 2\varepsilon\ll E_g$. With weak interaction, an exciton is a well-defined \emph{excited} state, so that the system does not undergo an excitonic phase transition. In addition, with weak interaction, the dispersive bands are negligible when we consider excitations with low energy $\lesssim 2\varepsilon$. Indeed, perturbation theory suggests that there is a relative correction of order $\mathcal{O}(U/E_g)\ll 1$ from dispersive band. This is the reason that we neglected fermionic operators of the dispersive bands in the electron-hole state and the contact interaction, Eqs.~(\ref{Eq:GeneralP-HState}) and (\ref{Eq:ContactInteraction}).
	
	Now, we are ready to write down the Schrodinger equation:
	\begin{equation}
		(E-2\epsilon)|e_1,h_2;\boldsymbol{Q}\rangle =H_{\text{int}}|e_1,h_2;\boldsymbol{Q}\rangle
		\label{Eq:SchrodingerEq}
	\end{equation}
	where $2\epsilon$ is the energy needed to excite an electron-hole pair in a \emph{noninteracting} system, Fig.~\ref{fig:LiebLattice_Exciton_Setup}. This equation can be solved quite straightforwardly numerically and in some cases analytically. In particular, it captures the exciton excitations below the electron-hole continuum. We will show that the exciton's effective mass and its wavefunction spread are both determined by the quantum metric. More details are as follows.
	
	\subsection{Exciton's effective mass}

	\begin{figure}[t]
		\centering
		\includegraphics[width = 0.5\textwidth]{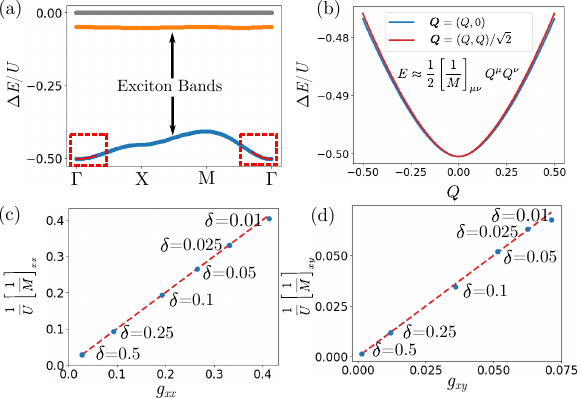}
		\caption{(a) Lowest three branches of electron-hole excitations for Lieb lattice hopping model at $\delta=0.5$. The lower two branches (blue and orange) are the exciton bands, which are singly degenerate. The top branch (gray) is the dense electron-hole ``continuum'', which collapses to a line due to flat bands. (b) Zoom in of the lower exciton band around $\Gamma$-point (red dashed box in (a)), which has an effective mass description $E\approx \frac{1}{2}\left[\frac{1}{M}\right]_{\mu\nu}Q^{\mu}Q^{\nu}$. There is small difference in the exciton band along two paths in BZ: $\boldsymbol{Q}=(Q,0)$ (blue) and $\boldsymbol{Q}=(Q,Q)/\sqrt{2}$ (red). (To exaggerate the difference, the dispersion is computed at $\delta=0.01$.) (c-d)Exciton effective mass (blue dots) at various choices of Lieb lattice parameter $\delta$. The effective mass is compared with the quantum metric $g_{xx}$ and $g_{xy}$(red dashed line) at the corresponding parameter $\delta$. The effective mass is measured in unit of interaction strength $U$.}
		\label{fig:LiebBilayer_Exciton_Band&Mass}
	\end{figure}
	
	Fig.~\ref{fig:LiebBilayer_Exciton_Band&Mass}(a) shows the spectra of electron-hole excitations in the Lieb lattice bilayer system at low energy. There are three ``bands'' of electron-hole excitations. The top ``band'' (gray) corresponds to the dense unbounded electron-hole excitations, which is typically referred to as particle-hole ``continuum'' (though, the ``continuum'' collapses to a line in flat band system under consideration).
	
	More importantly is the two exciton bands below the particle-hole ``continuum'' (blue and orange bands in Fig.~\ref{fig:LiebBilayer_Exciton_Band&Mass}(a)). It is obvious that the lower exciton band (blue) is dispersive. In particular, near $\Gamma$-point ($\boldsymbol{Q}=0$), the exciton kinetic energy has an effective mass description $E=\frac{1}{2}\left[\frac{1}{M}\right]_{\mu\nu}Q^{\mu}Q^{\nu}$. Notice that the constituent electron and hole are both extremely heavy, meanwhile the composite object, namely the excitons, are much lighter.  This counter-intuitive result has a quantum mechanical origin.
	
	A perturbative caluclation shows that the inverse of the effective mass tensor is proportional to the quantum metric (see Appendix.~\ref{sec:SchrodingerEqaution&GaugeInvariance}):
	\begin{equation}
		\frac{1}{U}\left[\frac{1}{M}\right]_{\mu\nu}=g_{\mu\nu}.
	\end{equation}
	This result is confirmed by numerical calculation, Fig.~\ref{fig:LiebBilayer_Exciton_Band&Mass}(b)-(d). Fig.~\ref{fig:LiebBilayer_Exciton_Band&Mass}(b) zooms in the exciton dispersion around $\Gamma$-point along different trajectories in momentum space, namely $\boldsymbol{Q}=(Q,0)$ and $\boldsymbol{Q}=(Q,Q)/\sqrt{2}$. The slight difference of the dispersions in Fig.~\ref{fig:LiebBilayer_Exciton_Band&Mass}(b) allows us to extract the full components of the inverse effective mass tensor. Fig.~\ref{fig:LiebBilayer_Exciton_Band&Mass}(c)-(d) shows that $M^{-1}_{xx}$ and $M^{-1}_{xy}$ shows good agreement with the quantum metric at various parameter choices $\delta$ of the Lieb lattice.
	
	Qualitatively, one may argue that the effective mass of excitons is closely related to the spatial quantum flutuactions as dictated by the quantum metric. Indeed, the wavefunction overlap of electron and hole is generally dependent on the total momentum $\boldsymbol{Q}$, leading to a different binding energy for excitons with different momenta. If there were vanishing quantum metric, the excitons will be fully localized with infinite effective mass.

	\subsection{Exciton size}
	
	\begin{figure}[t]
	\centering
	\includegraphics[scale = 1.25]{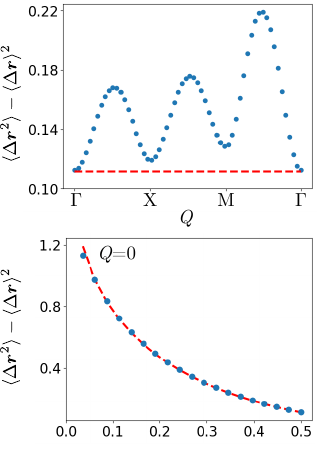}
	\caption{Flat band exciton size vs: (a) momentum $\boldsymbol{Q}$ at $\delta=0.5$; (b) $\delta$ at $\Gamma$-point ($\boldsymbol{Q}=0$). The red dashed line is twice of the quantum metric, $2\text{tr}\left(g_{\mu\nu}\right)$, at corresponding values of $\delta$. }
	\label{fig:LiebBilayer_Exciton_Size&QM}
	\end{figure}
	
	As mentioned, quantum metric describes the spatial spread the wavefunctions. In this subsection, we will show the exciton size is also closely related to the quantum metric. We define the size of exciton through its real space wavefunction. Direct calculation shows that the average separation between electron and hole is quite small. Therefore, we define the exciton size as the variance of the electron-hole separation.
	
	To define the size of excitons, we need to look at the wavefunction, Eq.~(\ref{Eq:GeneralP-HState}), more carefully. We take the exciton wavefunction at $\boldsymbol{Q}=0$ in the lower exciton band as an example. The exciton wavefunction at $\boldsymbol{Q}=0$ in the lower exciton band can be solved exactly as:
	\begin{equation}
		|e_1,h_2;\boldsymbol{Q}=0 \rangle  =\int_{\text{B.Z.}}\frac{d^2k}{(2\pi)^2}c^{\dagger}_{\boldsymbol{k}}d_{\boldsymbol{k}}|\text{G.S.}\rangle
		\label{Eq:EXWF_Q=0}
	\end{equation}
	It straightforward to rewrite this wavefunction in terms of the fermionic operators in real space:
	\begin{equation}
	\begin{split}
			|e_1,h_2;\boldsymbol{Q}=0\rangle =&\sum_{\boldsymbol{r}_1,\boldsymbol{r}_2}\int_{\text{B.Z.}}\frac{d^2k}{(2\pi)^2}\sum_{\alpha,\beta=\text{A,B,C}}\\
			\times&\psi_{\boldsymbol{k},\alpha}(\boldsymbol{r}_1)\psi^*_{\boldsymbol{k},\beta}(\boldsymbol{r}_2)c^{\dagger}_{\boldsymbol{r}_1,\alpha}d_{\boldsymbol{r}_2, \beta}|\text{G.S.}\rangle
	\end{split}
	\end{equation}
	where $c^{\dagger}_{\boldsymbol{r},\alpha}$ ($d_{\boldsymbol{r},\alpha}$) is the electron creation (annihilation) operator at the sublattice $\alpha$ within unit cell at $\boldsymbol{r}$ in layer 1(2). Special attention should be paid to the envelop wavefunction:
	\begin{equation}
		\psi^{\text{env}}_{\alpha,\beta}(\boldsymbol{x})=\int_{\text{B.Z.}}\frac{d^2k}{(2\pi)^2}e^{i\boldsymbol{k}\cdot\boldsymbol{x}}u_{\boldsymbol{k},\alpha}u_{\boldsymbol{k},\beta}^*
		\label{Eq:EnvWF}
	\end{equation}
	where $\boldsymbol{x}=\boldsymbol{r}_1-\boldsymbol{r}_2$. This envelop wavefunction describes the probablity amplitude of finding a separation of $\boldsymbol{x}$ between the electron and the hole forming the exciton.
	
	Before calculating the exciton size, we should comment that the envelop wavefunction, Eq.~(\ref{Eq:EnvWF}), is gauge invariant. To see this point, we need to emphasize that there is a hidden function $\psi(\boldsymbol{k},\boldsymbol{Q}=0)=1$, originating from the construction of the exciton wavefunction Eq.~(\ref{Eq:GeneralP-HState}). The result of $\psi(\boldsymbol{k},\boldsymbol{Q}=0)=1$ follows from assuming the same gauge choice for the Bloch wavefunctions of both Lieb lattice layers. Indeed, the gauge invariance of the envelop wavefunction can be deduced from the Schrodinger equation, Eq.~(\ref{Eq:SchrodingerEq}) (see Appendix.~\ref{sec:SchrodingerEqaution&GaugeInvariance}).
	
	Straightforward calculation shows that the average separation between the electron and hole vanishes exactly for the exciton described by the wavefunction, Eq.~(\ref{Eq:EXWF_Q=0}):
	\begin{equation}
		\langle \boldsymbol{x}\rangle_{\boldsymbol{Q}=0}=\sum_{\boldsymbol{x}}\sum_{\alpha,\beta}\boldsymbol{x}\left|\psi^{\text{env}}_{\alpha,\beta}(\boldsymbol{x})\right|^2=0
	\end{equation}
	For the model under consideration, the average separation between the electron and the hole remains small compared to the lattice constant and to the variance of electron-hole separation.
	
	Instead, we define the exciton size as the quadratic spread (variance) of the exciton wavefunction as:
	\begin{equation}
		\langle \Delta\boldsymbol{x}^2\rangle_{\boldsymbol{Q}=0}=\sum_{\boldsymbol{x}}\sum_{\alpha,\beta}\boldsymbol{x}^2\left|\psi^{\text{env}}_{\alpha,\beta}(\boldsymbol{x})\right|^2-\langle\boldsymbol{x}\rangle^2_{\boldsymbol{Q}=0}
	\end{equation}
	Again, for the wavefunction of Eq.~(\ref{Eq:EXWF_Q=0}), the calculation is quite straightforward. One find that for the exciton wavefunction of Eq.~(\ref{Eq:EXWF_Q=0}), the quadratic spread is given by twice of the trace of quantum metric tensor:
	\begin{equation}
		\langle \Delta\boldsymbol{x}^2\rangle_{\boldsymbol{Q}=0}=2\ \text{tr}\left(g_{\mu\nu}\right)
		\label{Eq:EXSize_Q=0}
	\end{equation}
	Indeed, the size of a flat band exciton can be determined by the quantum metric. More generally, the size of an exciton can be calculated numerically. The result is summarized in Fig.~\ref{fig:LiebBilayer_Exciton_Size&QM}.
	
	Fig.~\ref{fig:LiebBilayer_Exciton_Size&QM}(a) shows the size of an exciton with various total momentum $\boldsymbol{Q}$ at a parameter choice $\delta=0.5$. We find that the generally, the exciton size is larger than the value set by twice of the quantum metric (indicated by the red dashed line). In addition, we find that for the model under consideration, the exciton with total momentum $\boldsymbol{Q}=0$ has the minimal size, saturating to the value predicted by Eq.~(\ref{Eq:EXSize_Q=0}), namely the quantum metric.
	
	Fig.~\ref{fig:LiebBilayer_Exciton_Size&QM}(b) shows the numerical calculation of the size of excitons with zero total momentum $\boldsymbol{Q}=0$ at various values of Lieb lattice model parameter $\delta$. The computed exciton sizes (blue dots) are compared with the quantum metric prediction, Eq.~(\ref{Eq:EXSize_Q=0}) (red dashed line). There is a good match between the numerical caluclation and the value set by the quantum metric. 
	
	We should mention that in Fig.~\ref{fig:LiebBilayer_Exciton_Size&QM}(b), there is a deviation between the numerical data and the prediction of Eq.~(\ref{Eq:EXSize_Q=0}) at small $\delta$. The deviation is related to the detailed distribution of quantum metric in Brillouin zone. Indeed, for small $\delta$, the quantum metric is sharply peaked at momentum $\boldsymbol{k}=(\pi,\pi)$, which makes accurate numerical calculation time consuming.
	
	To conclude, the size of flat band excitons is lower bounded by the quantum metric.
	
	\section{Implications on exciton condensate}\label{sec:EXCondensate}
	
	%The discussion of a single exciton's kinetic energy and wavefunctions has direct implications on the many body physics, in particular the exciton condensate. The complication comes from the unavoidable repulsive interaction between excitons.
	
	In the previous sections, we discussed the properties of a single exciton excitation above an insulating groundstate in a bilayer system. Importantly, due to quantum metric, the flat band excitons acquire a finite kinetic energy, as dictated by the effective mass, and a finite wavefunction extension, which in principle can grow indefinitely by the quantum metric. The kinetic energy and wavefunction extension of flat band excitons imply the possibility of forming exciton condensate.
	
	In the absence of repulsive interaction between excitons, the kinetic energy always dominates. Therefore, excitons can condense to a superfluid phase in this ideal situation. The reason behind is the quantum metric, which gives flat band excitons an effective mass, namely kinetic energy. At the meantime, the spatial quantum fluctuations can be coupled nonlocally due to the quantum metric and the extended wavefunction.  Phase of the superfluid order parameter can be synchronized over space. Thus, a phase coherent state is possible. 
	
	In reality, the repulsive interaction between excitons is typically inevitable, for example dipole-dipole interactions. However, previous studies revealed that the interaction energy can be significantly reduced and even bounded at short distance \cite{PhysRevB.84.075130,filinov2009effective,zimmermann2007exciton}. This is because the finite spatial extension of exciton's wavefunction. Therefore, depending on the details, the repulsive interaction between excitons may still be weak, compared to typical kinetic energy. Thus, exciton condensate is indeed possible in practical situations. Again, for flat band systems, quantum metric is the key issue behind, which affects the comparison of kinetic energy and repulsive energy of excitons \cite{PhysRevB.84.075130,PhysRevB.76.064511}.
	
	Despite the complications in realistic models, we focus on the physics deep in the exciton condensate phase, where the role of quantum metric can be revealed explicitly \cite{PhysRevLett.127.170404,PhysRevB.101.060505,PhysRevLett.132.026002}. 
	
\begin{figure}[t]
		\centering
		\includegraphics[scale = 1.25]{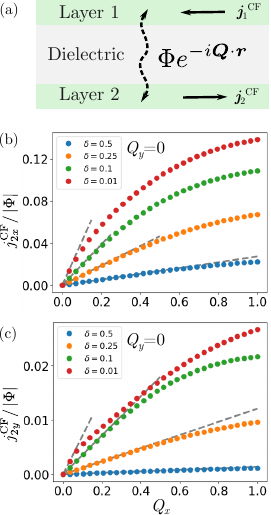}
		\caption{(a) Condensation of indirect exciton is described by an inter-layer coherence order parameter $\Phi e^{-i\boldsymbol{Q}\cdot\boldsymbol{x}}$. The non-zero phase gradient $\boldsymbol{Q}$ gives rise to a non-dissipative counterflow (CF) supercurrent $\boldsymbol{j}_{1,2}^{\text{CF}}$ in the bilayer system. (b-c) CF current of layer 2 vs the phase gradient $\boldsymbol{Q}=(Q_x,0)$ at balanced filling $\nu_1=\nu_2=1/2$ (namely, the chemical potentials are $\mu_1=\mu_2=0$) at zero temperature. The CF supercurrent at small $\boldsymbol{Q}$ is compared with the analytical result $j_{2,\mu}^{\text{CF}}=|\Phi|g_{\mu\nu}Q^{\nu}$  (gray dashed line).}
		\label{fig:CFCurrent_vs_Q}
	\end{figure}

	One immediate result is that the exciton condensate can carry a \emph{non-dissipative} counterflow (CF) supercurrent, Fig.~\ref{fig:CFCurrent_vs_Q}(a) \cite{su2008make,nandi2012exciton,nguyen2023perfect}. Instead of a voltage bias, such non-dissipative CF supercurrent can be captured by an inter-layer coherence order parameter. As shown in Fig.~\ref{fig:CFCurrent_vs_Q}(a), the order parameter, $\Phi e^{-i\boldsymbol{Q}\cdot\boldsymbol{r}}$, acquires a phase gradient. The CF supercurrent is proportional to the phase gradient, $j^{\text{CF}}_{\mu}\propto g_{\mu\nu}Q^{\nu}$, with the proportionality given by the quantum metric, Fig.~\ref{fig:CFCurrent_vs_Q}.
	
	To see this point, we can consider the following mean field model for bilayer of Lieb lattice hopping model. Namely, the mean field order parameter introduces an inter-layer coherent term to the Hamiltonian: $\delta H=\sum_{\boldsymbol{r}}\sum_{\alpha} e^{-i\boldsymbol{Q}\cdot\boldsymbol{r}}\Phi_{\alpha}c^{\dagger}_{\alpha}(\boldsymbol{r})d_{\alpha}(\boldsymbol{r})+\text{h.c.}$. In momentum space, the Hamiltonian takes a block diagonal form as follows:
	\begin{equation}
		\begin{split}
			& H_{\text{MF}}(\boldsymbol{k})=\begin{bmatrix}
			H(\boldsymbol{k})-\mu_1 & \Phi_{\text{mat}}\\
			\Phi_{\text{mat}}^{\dagger} & H(\boldsymbol{k}-\boldsymbol{Q})-\mu_2
		\end{bmatrix}\\
		&\Phi_{\text{mat}}=\text{Diag}[\Phi,0,\Phi]
		\end{split}
		\label{Eq:MF_Ham}
	\end{equation}
	where the diagonal terms are the Bloch Hamiltonian of Lieb lattice hopping model, Eq.~(\ref{Eq:LiebLattice_BlochH}), for layer 1 and 2 respectively; $\Phi_{\text{mat}}$ is the order parameter that coherently couples two Lieb lattice layers; $\mu_{1,2}$ are the chemical potentials, which defines the filling factor of two layers. 
	
	We consider chemical potentials being close to the flat band, with total filling $\nu_{\text{tot}}=1$ (excluding filling from the lower dispersive bands, Fig.~\ref{fig:LiebLattice_LatticeStructure&BandStructure&QM}). In this case, the chemical potential has the following relation: $\mu_1=-\mu_2=\mu$. Now, the flat bands are strongly coupled by the order parameter $\Phi$. The quasiparticle excitation has the following spectrum:
	\begin{equation}
		\epsilon_{\pm}(\boldsymbol{k})=\pm\epsilon(\boldsymbol{k})=\pm\sqrt{\mu^2+\left|u^{\dagger}_{\boldsymbol{k}-\boldsymbol{Q}}\Phi_{\text{mat}}u_{\boldsymbol{k}}\right|^2}
	\end{equation}
	The filling of each layer can be determined through the following equation $2\nu-1=\mu\int_{\text{BZ}}\frac{d^2k}{(2\pi)^2}\frac{\tanh \left[\beta\epsilon(\boldsymbol{k})/2\right]}{\epsilon(\boldsymbol{k})}$
	with the filling of layer 1 being $\nu_1=\nu$ and layer 2 $\nu_2=1-\nu$; $\beta=1/T$ is inverse of temperature.
	
	Now, we are ready to write down the expression for the current. The full expression is a bit cumbersome, Eq.~(\ref{Eq:CFCurrent_FullExpression}). Nevertheless, for small phase gradient $|\boldsymbol{Q}|\ll 1$, quantum metric shows up explicitly in the CF supercurrent:
	\begin{equation}
		j_{2,\mu}^{\text{CF}}=-j_{1,\mu}^{\text{CF}}=|\Phi|^2\int_{\text{BZ}}\frac{d^2k}{(2\pi)^2}\frac{\tanh \left[\beta\epsilon(\boldsymbol{k})/2\right]}{\epsilon(\boldsymbol{k})}g_{\mu\nu}(\boldsymbol{k})Q^{\nu}
		\label{Eq:CFCurrent}
	\end{equation}
	The CF supercurrent in layer 1 is exactly opposite to that in layer 2. Fig.~\ref{fig:CFCurrent_vs_Q}(b-c) plots the CF supercurrent in layer 2 for a phase gradient $\boldsymbol{Q}=(Q_x,0)$ at half filling for both layers at zero temperature. At half filling, the chemical potential is zero $\mu=0$. The expression for the CF supercurrent is simplified to $j_{2,\mu}^{\text{CF}}=|\Phi|g_{\mu\nu}Q^{\nu}$ at small $\boldsymbol{Q}$. This simplified expression is plotted as the gray dashed line in Fig.~\ref{fig:CFCurrent_vs_Q}(b-c).
	
	Meanwhile, the solid dots in Fig.~\ref{fig:CFCurrent_vs_Q}(b-c) plots the CF supercurrent based on the full expression, Eq.~(\ref{Eq:CFCurrent_FullExpression}).  At small $Q_x$, the CF supercurrent is linear in phase gradient, matching the analytical result of Eq.~(\ref{Eq:CFCurrent}) (gray dashed line in Fig.~\ref{fig:CFCurrent_vs_Q}(b-c)). Indeed, the quantum metric plays a key role in exciton condensate. Namely, without quantum metric, dissipationless CF current is not possible.
	
	We should mention that Eq.~(\ref{Eq:CFCurrent}) is not most general form for CF supercurrent. Indeed, Eq.~(\ref{Eq:CFCurrent}) follows from the specific form of order parameter, Eq.~(\ref{Eq:MF_Ham}), and the structure of Bloch wavefunction, Eq.~(\ref{Eq:FB_BlochWF}). One can perform the analysis in a fully self-consistent manner. In the end, the CF current may deviate from the form of Eq.~(\ref{Eq:CFCurrent}). Nevertheless, the contribution from qunatum metric, namely Eq.~(\ref{Eq:CFCurrent}), is not negligible \cite{PhysRevB.106.014518}.
	
	The significance of dissipationless CF supercurrent can be seen from two perspective. First of all, the dissipationless CF supercurrent takes place in both topological and non-topological flat bands. The former corresponds to quantum Hall systems.The composite fermions acquires a finite kineitc energy in the composite Fermi liquid description \cite{PhysRevB.95.085135,doi:10.1146/annurev-conmatphys-033117-054227}. In contrast, the non-topological flat band corresponds to systems like Lieb lattice hopping model, where the electrons or holes are extremely heavy. However, because of quantum metric, (non-dissipative) transport is possible at zero temperature. 
	
	Second, the quantum metric stablizes the superfluid phase. Indeed, the presence of CF current, Eq.~(\ref{Eq:CFCurrent}), implies a nonzero superfluid phase stiffness \cite{PhysRevB.101.060505}. Qualitatively, finite superfluid phase stiffness supresses the phase fluctuations and thus protects the phase coherence. 
	
	\section{Conclusion and outlook} \label{sec: discussion}
	
	To conclude, we investigate the properties of flat band excitons, with an emphasis on the quantum metric. We found that due to the quantum metric, the properties of excitons is drastically different from the conventional expectation based on hydrogen model. The mechanism behind is rooted in the spatial quantum fluctuations of the wavefunctions, as dictated by the quantum metric. Several result follows:
	
	First, the excitons acquire a finite effective mass, the inverse of which was shown to be proportional to the quantum metric. In addition, the excitons are not point-like particles. Instead, they have a finite size, which is lower bounded by the quantum metric.
	
	Second, quantum metric aids stablizing the exciton condensate and makes transport possible through flat band system. Indeed, we found that a dissipationless CF supercurrent can be established in the condensation of spatially indirect excitons. Transport implies local fluctuation in particle number and suppresses the phase fluctuation. Hence, a phase coherent condensate is stable. Notice that here topology is not required. Only nonvanishing quantum metric (or equivalently the spatial quantum fluctuation) is required.
	
	We demonstrated the results with a hopping model defined on the Lieb lattice. The Lieb lattice can be realized by cold atoms in optical lattices \cite{PhysRevB.81.041410,taie2015coherent} as well as carefully designed electronic lattice \cite{slot2017experimental}.  Indeed, the Bose-Einstein condensate (BEC) of exciton-polariton was observed in 1D Lieb lattice \cite{PhysRevLett.116.066402}. In fact, the results reported in this article is not limited to Lieb lattice. Lieb lattice is one representative example of a wide range of flat band systems \cite{10.21468/SciPostPhys.15.4.139,PhysRevB.99.045107,cualuguaru2022general,regnault2022catalogue}.
	
	As pointed out by Ref.~\cite{PhysRevLett.116.066402}, the flat band BEC is very sensitive to disorder. Localized states show up in the presence of disorder. Another competing effect of flat band BEC is the Wigner crystallization \cite{PhysRevLett.99.070401,liu2022crossover}. Whether quantum metric can drive a localization-delocalization transition or quantum melting of Wigner crystal can be an interesting direction for future work.

	\acknowledgements
	We are grateful to Shuai Chen, and Kangle Li for discussions. K. T. L.
	acknowledges the support of the Ministry of Science and Technology, China and the Hong Kong Research Grants Council through Grants No. 2020YFA0309600, No. RFS2021-6S03, No. C6025-19G, No. AoE/P701/20, No. 16310520, No. 16310219, No. 16307622, and No. 16309223.
	
	\appendix

	\section{Exciton real space wavefunction in quantum Hall bilayer} \label{sec:LLEX_RealSpaceWF_ExplicitForm}
	
	In this section, we present the excplicit form of the real space exciton wavefunction in quantum Hall bilayer, Eq.~(\ref{Eq:LLEX_RealSpaceWF}), as follows:
	\begin{equation}
		\begin{split}
			\psi^{\boldsymbol{Q}}_{\text{AA}}(\boldsymbol{r}_1,\boldsymbol{r}_2)=&\mathcal{N}e^{iQ_y\bar{y}}e^{i\frac{Q_y}{2}\Delta y}e^{i\frac{B}{2}(2\bar{x}-\frac{Q_y}{B})(\Delta y + \frac{Q_x}{B})}\\
			\times&\left(\frac{B}{2\pi}\right)^2e^{-\frac{B}{4}(\Delta x-\frac{Q_y}{B})^2}e^{-\frac{B}{4}(\Delta y-\frac{Q_x}{B})^2}
		\end{split}
	\end{equation}
	where on the left hand side $\boldsymbol{r}_{1,2}$ are the locations of the electron and hole respectively; on the right hand side, we rewrite the spatial coodinates as the average position and relative position of the electron and hole: $(\bar{x},\bar{y})=(\boldsymbol{r}_1+\boldsymbol{r}_2)/2$ and $(\Delta x, \Delta y)=\boldsymbol{r}_1-\boldsymbol{r}_2$.
	
	Given the explicit form of the exciton's real space wavefunction, we are able to read out the probability of finding the electron and hole at a separation of $\Delta \boldsymbol{r}=(\Delta x, \Delta y)$:
	\begin{equation}
		\rho(\Delta \boldsymbol{r})=\frac{B}{2\pi}e^{-\frac{B}{2}(\Delta x-\frac{Q_y}{B})^2}e^{-\frac{B}{2}(\Delta y-\frac{Q_x}{B})^2}
	\end{equation}
	The average and variance of the separation between electron and hole in the main text follows from the probability distribution above. Similar analysis can be carried out for general flat band excitons.

	\section{Explicit form of Schrodinger equation Eq.~(\ref{Eq:SchrodingerEq}) and its solution}
	
	\label{sec:SchrodingerEqaution&GaugeInvariance}
	
	In this section, we provide the explicit equation for the function $\psi(\boldsymbol{k},\boldsymbol{Q})$ in determining the exciton wavefunction, Eq.(\ref{Eq:GeneralP-HState}). The equation for $\psi(\boldsymbol{k},\boldsymbol{Q})$ follows directly from the Schodinger equation Eq.~(\ref{Eq:SchrodingerEq}):
	\begin{equation}
		\begin{split}
			\psi(\boldsymbol{k},\boldsymbol{Q})=-&\frac{U}{E-E_{\text{GS}}-2\epsilon}\sum_{\alpha}u_{\boldsymbol{k}+\boldsymbol{Q},\alpha}^*u_{\boldsymbol{k},\alpha}\\
			\times&\int\frac{d^2q}{(2\pi)^2}u_{\boldsymbol{q}+\boldsymbol{Q},\alpha}u_{\boldsymbol{q},\alpha}^*\ \psi(\boldsymbol{q},\boldsymbol{Q})
		\end{split}
		\label{Eq:Explicit_SchrodingerEq}
	\end{equation}
	From this equation, we may argue for the gauge invariance of the exciton's real space wavefunction for an exciton with total momentum $\boldsymbol{Q}$:
	\begin{equation}
			\psi^{\text{env},\boldsymbol{Q}}_{\alpha,\beta}(\boldsymbol{x})=\int_{\text{B.Z.}}\frac{d^2k}{(2\pi)^2}e^{i\boldsymbol{k}\cdot\boldsymbol{x}}u_{\boldsymbol{k}+\boldsymbol{Q},\alpha}u_{\boldsymbol{k},\beta}^*\ \psi(\boldsymbol{k},\boldsymbol{Q}).
			\label{Eq:FBEX_RealSpaceWF}
	\end{equation}
	Indeed, let us make a gauge transformation on the periodic part of the Bloch wavefunctions,
	\begin{equation}
		u_{\boldsymbol{k},\alpha}\rightarrow e^{i\phi(\boldsymbol{k})}u_{\boldsymbol{k},\alpha}
	\end{equation}
	where the phase $\phi(\boldsymbol{k})$ is the same for all components in $u_{\boldsymbol{k}}$. Then, the Schrodinger equation, Eq.(\ref{Eq:Explicit_SchrodingerEq}), suggests the following tranformation on $\psi(\boldsymbol{k},\boldsymbol{Q})$:
	\begin{equation}
		\psi(\boldsymbol{k},\boldsymbol{Q})\rightarrow e^{i\phi(\boldsymbol{k})-i\phi(\boldsymbol{k}+\boldsymbol{Q})}\psi(\boldsymbol{k},\boldsymbol{Q}).
	\end{equation}
	It's straightforward to see that the wavefunction, Eq.~(\ref{Eq:FBEX_RealSpaceWF}), is invariant under the gauge transformation discussed above. In Eq.~(\ref{Eq:EnvWF}) in the maintext, $\psi(\boldsymbol{k},\boldsymbol{Q}=0)=1$ is a result of the a particular gauge choice as discussed in the paragraph below Eq.~(\ref{Eq:EnvWF}).
	
	To solve the Schrodinger equation, one can discretize BZ and diagonalize Eq.~(\ref{Eq:Explicit_SchrodingerEq}). The result is the three branches of electron-hole excitations in Fig.~\ref{fig:LiebBilayer_Exciton_Band&Mass}(a). If we focus on the exciton bands, we can multiply $\int_{\text{BZ}}\frac{d^2k}{(2\pi)^2}u^*_{\boldsymbol{k},\beta}u_{\boldsymbol{k}+\boldsymbol{Q},\beta}\times$ Eq.~(\ref{Eq:Explicit_SchrodingerEq}) and define:
	\begin{equation}
		\Psi_{\alpha}(\boldsymbol{Q})=\int_{\text{BZ}}\frac{d^2k}{(2\pi)^2}u^*_{\boldsymbol{k},\alpha}u_{\boldsymbol{k}+\boldsymbol{Q},\alpha}\psi(\boldsymbol{k},\boldsymbol{Q})
	\end{equation}
	and
	\begin{equation}
		\mathbb{U}_{\alpha\beta}(\boldsymbol{Q})=\int_{\text{BZ}}\frac{d^2k}{(2\pi)^2}u^*_{\boldsymbol{k},\alpha}u_{\boldsymbol{k}+\boldsymbol{Q},\alpha}u^*_{\boldsymbol{k}+\boldsymbol{Q},\beta}u_{\boldsymbol{k},\beta}
	\end{equation}
	Then, the Schrodinger equation, Eq.~(\ref{Eq:Explicit_SchrodingerEq}), reduces to:
	\begin{equation}
		-\frac{E-E_{\text{GS}}-2\epsilon}{U}\Psi_{\alpha}(\boldsymbol{Q})=\sum_{\beta}\mathbb{U}_{\alpha\beta}(\boldsymbol{Q})\Psi_{\beta}(\boldsymbol{Q})
	\end{equation}
	which is now a much simpler matrix diagonalization problem. 
	
	Focus on the lower exciton band, it's straightforward to find that the binding energy at $\boldsymbol{Q}=0$  is given by $E_0-E_{\text{GS}}-2\epsilon=-U/2$. In addtion, we find $\Psi_{\text{A}}(0)=\Psi_{\text{C}}(0)=1/\sqrt{2}$ and $\Psi_{\text{B}}(0)=0$.
	
	At small $\boldsymbol{Q}$, we try to find the binding energy perturbatively $E\approx E_0+\delta E$. Standard perturbation theory suggests that $\delta E=-U\sum_{\alpha\beta}\Psi^*_{\alpha}(0)\left[\mathbb{U}_{\alpha\beta}(\boldsymbol{Q})-\mathbb{U}_{\alpha\beta}(0)\right]\Psi_{\beta}(0)$. More explicitly, we can find the following expression:
	\begin{equation}
		\begin{split}
			\delta E=&-\frac{U}{2}\int_{\text{BZ}}\frac{d^2k}{(2\pi)^2}\left[\left|u^{\dagger}_{\boldsymbol{k}+\boldsymbol{Q}}u_{\boldsymbol{k}}\right|^2 -1\right]\\
			\approx&\frac{1}{2}Ug_{\mu\nu}Q^{\mu}Q^{\nu}
		\end{split}
	\end{equation}
	The second line follows from the definition of quantum metric. Indeed, the inverse effective mass tensor is given by the quantum metric.
	
	\section{Mean field analysis of exciton condensate} 
	\label{sec: MF_EXCondensate}
	
	In this section, we provide details in deriving the dissipationless CF supercurrent, Eq.~(\ref{Eq:CFCurrent}).
	
	To begin with, we rewrite the MF Hamiltonian:
	\begin{equation}
		\begin{split}
			& H_{\text{MF}}(\boldsymbol{k}-\boldsymbol{A})=\begin{bmatrix}
				H(\boldsymbol{k}-\boldsymbol{A}_1)-\mu & \Phi_{\text{mat}}\\
				\Phi_{\text{mat}}^{\dagger} & H(\boldsymbol{k}-\boldsymbol{Q}-\boldsymbol{A}_2)+\mu
			\end{bmatrix}
		\end{split}
		\label{Eq:MF_Ham_app}
	\end{equation}
	where the order parameter $\Phi_{\text{mat}}$ can take a more complicated matrix form, coupling different sublattices/orbitals of the two layers. In addition, we couple the MF Hamiltonian to two auxilliary gauge fields $\boldsymbol{A}_{1,2}$, which will aid extracting currents in both layers.
	
	It's standard exercise to write down partition function formally as:
	\begin{equation}
		\mathcal{Z}_0[\boldsymbol{A}]=\int D\bar{\psi}D\psi e^{\int_{0}^{\frac{1}{T}} d\tau \int\frac{d^2k}{(2\pi)^2}\bar{\psi}(\boldsymbol{k},\tau)\left[-\partial_{\tau}-H_{\text{MF}}(\boldsymbol{k}-\boldsymbol{A})\right]\psi(\boldsymbol{k},\tau)}
	\end{equation}
	and free energy as:
	\begin{equation}
		\mathcal{F}_0[\boldsymbol{A}]=-T\ln \mathcal{Z}_0[\boldsymbol{A}].
	\end{equation}
	The thermal dynamical quantities can be defined through derivatives on the free energy. In particular, the current can be evaluated as: 
	\begin{equation}
		\boldsymbol{j}^{\text{CF}}_l=-\frac{1}{T}\left.\frac{\delta\mathcal{F}_0[\boldsymbol{A}]}{\delta\boldsymbol{A}_l}\right|_{\boldsymbol{A}_1=\boldsymbol{A}_2=0}
	\end{equation}
	where $l=1,2$ is the layer index. More explicitly, the CF current can be computed as:
	\begin{equation}
		\boldsymbol{j}^{\text{CF}}_l=-T\sum_{i\omega_m}\int\frac{d^2k}{(2\pi)^2}\text{tr}\ G_0(i\omega_m,\boldsymbol{k})\ \hat{\boldsymbol{j}}_l(\boldsymbol{k})
	\end{equation}
	where the Green's function is defined as $G_0^{-1}(i\omega_m,\boldsymbol{k})=i\omega_m-H_{\text{MF}}(\boldsymbol{k})$; the current operator is:
	\begin{equation}
		\begin{split}
			\hat{\boldsymbol{j}}_1=-\begin{bmatrix}
				\partial_{\boldsymbol{k}}H(\boldsymbol{k}) & 0\\
				0 & 0
			\end{bmatrix},\ 
			\hat{\boldsymbol{j}}_2=-\begin{bmatrix}
				 0 & 0\\
				0 & \partial_{\boldsymbol{k}}H(\boldsymbol{k}-\boldsymbol{Q})
			\end{bmatrix}
		\end{split}.
	\end{equation}
	Notice that here we need to retain the full structure of the Bloch Hamiltonian, instead of focusing only on the flat band. In the discussion of all previous sections, only flat band information is needed so that band index was not written explicitly. From now on, we recover the band indices to avoid confusions.
	
	We focus on deriving $\boldsymbol{j}^{\text{CF}}_1$. The CF current in layer 2, $\boldsymbol{j}^{\text{CF}}_2$, can be computed in the same manner. To proceed, we need to look at the structure of the Green's function and current operator more carefully. We rewrite Green's function and current operator in the \emph{Bloch band} basis of the Bloch Hamiltonian, $H(\boldsymbol{k})$.
	
	First, the inverse Green's function is
	\begin{equation}
		G_0^{-1}(i\omega_m,\boldsymbol{k})=i\omega_m-\begin{bmatrix}
			H_{\text{f}} & \Phi_{\text{fd}}\\
			\Phi_{\text{fd}}^{\dagger} & H_{\text{d}}
		\end{bmatrix}
	\end{equation}
	where the MF Hamiltonian is written in the block form. The subscript ``$\text{f,d}$'' represent the Hilbert space of flat bands and dispersive bands, respectively. Each block is defined as follows:
	\begin{equation}
		H_{\text{f}}=\begin{bmatrix}
			-\mu & u_{\text{f},\boldsymbol{k}}^{\dagger}\Phi_{\text{mat}}u_{\text{f},\boldsymbol{k}-\boldsymbol{Q}} \\
			u_{\text{f},\boldsymbol{k}-\boldsymbol{Q}}^{\dagger}\Phi_{\text{mat}}^{\dagger}u_{\text{f},\boldsymbol{k}} & \mu 
		\end{bmatrix}
	\end{equation}
	\begin{equation}
		H_{\text{d}}=\begin{bmatrix}
			\epsilon_{\text{d}}(\boldsymbol{k})-\mu & u_{\text{d},\boldsymbol{k}}^{\dagger}\Phi_{\text{mat}}u_{\text{d},\boldsymbol{k}-\boldsymbol{Q}} \\
			u_{\text{d},\boldsymbol{k}-\boldsymbol{Q}}^{\dagger}\Phi_{\text{mat}}^{\dagger}u_{\text{d},\boldsymbol{k}} & \epsilon_{\text{d}}(\boldsymbol{k}-\boldsymbol{Q})+\mu 
		\end{bmatrix}
	\end{equation}
	\begin{equation}
		\Phi_{\text{fd}}=\begin{bmatrix}
			0 & u_{\text{f},\boldsymbol{k}}^{\dagger}\Phi_{\text{mat}}u_{\text{d},\boldsymbol{k}-\boldsymbol{Q}} \\
			u_{\text{f},\boldsymbol{k}-\boldsymbol{Q}}^{\dagger}\Phi_{\text{mat}}^{\dagger}u_{\text{d},\boldsymbol{k}} & 0
		\end{bmatrix}
	\end{equation}
	where we put back the band index ``f,d'' on the periodic part of Bloch wavefunction, $u_{\text{f(d)},\boldsymbol{k}}$. 
	
	To obtain the Green's function, we assume a large gap between flat bands and the dispersive bands: $|\epsilon_{\text{d}}(\boldsymbol{k})|\gg |\Phi_{\text{mat}}|$. To the order of $\mathcal{O}\left(\frac{1}{\epsilon_{\text{d}}(\boldsymbol{k})}\right)$, the Green's function reads (the dependence on frequency and momentum, $(i\omega_m,\boldsymbol{k})$ is assumed):
	\begin{equation}
		G_0=\begin{bmatrix}
			G_{\text{f}} + G_{\text{f}}\circ\Phi_{\text{fd}}\circ G_{\text{d}}\circ\Phi_{\text{fd}}^{\dagger}\circ G_{\text{f}} & -G_{\text{f}}\circ\Phi_{\text{fd}}\circ G_{\text{d}}\\
			-G_{\text{d}}\circ\Phi_{\text{fd}}^{\dagger}\circ G_{\text{f}} & G_{\text{d}}
		\end{bmatrix}
	\end{equation}
	where $G_{\text{f,d}}=\left[i\omega_m-H_{\text{f,d}}\right]^{-1}$. Importantly, the off-diagonal block is not negligible. Namely, the current operator will give a contribution on the order of $\mathcal{O}(\epsilon_{\text{d}}(\boldsymbol{k}))$, leading to an $\mathcal{O}(1)$ contribution to the current.
	
	Next, we should present the current operator in the band basis:
	\begin{equation}
		\hat{\boldsymbol{j}}_1=\begin{bmatrix}
			0 & \boldsymbol{j}_{1,\text{fd}}\\
			\boldsymbol{j}_{1,\text{fd}}^{\dagger} & \boldsymbol{j}_{1,\text{d}}
		\end{bmatrix}
	\end{equation}
	with
	\begin{equation}
		\boldsymbol{j}_{1,\text{fd}}=\begin{bmatrix}
			u^{\dagger}_{\text{f},\boldsymbol{k}}\partial_{\boldsymbol{k}}H(\boldsymbol{k})u_{\text{d},\boldsymbol{k}} & 0\\
			0 & 0
		\end{bmatrix},\ \boldsymbol{j}_{1,\text{d}}=\text{Diag}\left[\partial_{\boldsymbol{k}}\epsilon_{\text{d}}(\boldsymbol{k})\right].
	\end{equation}
	The diagonal elements are the group veolocities of each band. Notice that $\boldsymbol{j}_{1,\text{fd}}$ is on the order of $\mathcal{O}(\epsilon_{\text{d}}(\boldsymbol{k}))$. Indeed, one can rewrite $\boldsymbol{j}_{1,\text{fd}}$ as:
	\begin{equation}
		\boldsymbol{j}_{1,\text{fd}}=\begin{bmatrix}
			-\epsilon_{\text{d}}(\boldsymbol{k})\left(\partial_{\boldsymbol{k}}u^{\dagger}_{\text{f},\boldsymbol{k}}\right)u_{\text{d},\boldsymbol{k}} & 0\\
			0 & 0
		\end{bmatrix}.
	\end{equation}
	
	Combining the Green's function and current operator, we are able to derive the CF current as:
	\begin{equation}
		\boldsymbol{j}^{\text{CF}}_1=-T\sum_{i\omega_m}\int\frac{d^2k}{(2\pi)^2}\text{tr}\left[G_{\text{f}}\Phi_{\text{fd}}G_{\text{d}}\boldsymbol{j}_{1,\text{fd}}^{\dagger}+\text{h.c.}\right]
		\label{Eq:CF_current_Expression}
	\end{equation}
	where the contribution from dispersive band, $\sim \text{tr}\ G_{\text{d}}\ \boldsymbol{j}_{1,\text{d}}$, is negligible. This is because the presence of Fermi function after the summation over Matsubara frequency. At low temperature, $T\ll \epsilon_{\text{d}}(\boldsymbol{k})$, the dispersive bands are either almost empty or almost filled. Both cases have exponentially small contribution, $\sim e^{-|\epsilon_{\text{d}}(\boldsymbol{k})|/T}$, to the current.
	
	Focus on Eq.~(\ref{Eq:CF_current_Expression}). Notice that the remaining Green's functions have the following structure:
	\begin{equation}
		G_{\text{f}}=\sum_{s=\pm}\frac{1}{i\omega_m-\epsilon_{s}(\boldsymbol{k})}\frac{1}{2}\left(\mathbb{I}+s\frac{H_{\text{f}}}{|\epsilon_{s}(\boldsymbol{k})|}\right)
	\end{equation}
	\begin{equation}
		G_{\text{d}}\approx\frac{1}{i\omega_m-\epsilon_{\text{d}}(\boldsymbol{k})}.
	\end{equation}
	We are able to obtain the following result of the CF current:
	\begin{equation}
		\begin{split}
			\boldsymbol{j}^{\text{CF}}_1=&\int\frac{d^2k}{(2\pi)^2}\frac{n_{\text{F}}\left(\epsilon_{-}(\boldsymbol{k})\right)-n_{\text{F}}\left(\epsilon_{+}(\boldsymbol{k})\right)}{2|\epsilon_{+}(\boldsymbol{k})|}\\
			\times &u^{\dagger}_{\text{f},\boldsymbol{k}}\Phi_{\text{mat}}u_{\text{f},\boldsymbol{k}-\boldsymbol{Q}}u^{\dagger}_{\text{f},\boldsymbol{k}-\boldsymbol{Q}}\Phi^{\dagger}_{\text{mat}}u_{\text{d},\boldsymbol{k}}u^{\dagger}_{\text{d},\boldsymbol{k}}\left(\partial_{\boldsymbol{k}}u_{\text{f},\boldsymbol{k}}\right) + \text{h.c.}
		\end{split}
	\end{equation}
	Or equivalently, one can express the CF current in terms of flat band wavefunctions only:
	\begin{equation}
		\begin{split}
			\boldsymbol{j}^{\text{CF}}_1=&\int\frac{d^2k}{(2\pi)^2}\frac{n_{\text{F}}\left(\epsilon_{-}(\boldsymbol{k})\right)-n_{\text{F}}\left(\epsilon_{+}(\boldsymbol{k})\right)}{2|\epsilon_{+}(\boldsymbol{k})|}\\
			\times &u^{\dagger}_{\text{f},\boldsymbol{k}}\Phi_{\text{mat}}u_{\text{f},\boldsymbol{k}-\boldsymbol{Q}}u^{\dagger}_{\text{f},\boldsymbol{k}-\boldsymbol{Q}}\Phi^{\dagger}_{\text{mat}}(\mathbb{I}-u_{\text{f},\boldsymbol{k}}u^{\dagger}_{\text{f},\boldsymbol{k}})\left(\partial_{\boldsymbol{k}}u_{\text{f},\boldsymbol{k}}\right) + \text{h.c.}
		\end{split}
		\label{Eq:CFCurrent_FullExpression}
	\end{equation}
	Further assumption of the structure of the order parameter, $\Phi_{\text{mat}}$, and the structure of $u_{\text{f},\boldsymbol{k}}$, as in Eq.~(\ref{Eq:LiebLattice_BlochH}) and Eq.~(\ref{Eq:FB_BlochWF}), one can obtain the CF supercurrent in the small phase gradient limit as reported in the maintext, Eq.~(\ref{Eq:CFCurrent}).
	
	Lastly, we comment on the determination of the order parameter. Indeed, one can add to the free energy following term:
	\begin{equation}
		\delta\mathcal{F}=\frac{1}{U}\sum_{\boldsymbol{r}}\sum_{\alpha}\left|\Phi_{\alpha}\right|^2
	\end{equation}
	which can be derived from the Hubbard-Stratonovich transformation of the contact interaction Eq.~(\ref{Eq:ContactInteraction}). The saddle point equation of the order parameter $\Phi_{\alpha}$ based on the free energy $\mathcal{F}_0+\delta\mathcal{F}$ determines the order parameter self-consistently.
	
	\bibliography{EX_QM_Ref}{}
	
\end{document}